\documentclass[prl,twocolumn,preprintnumbers,amsmath,amssymb,superscriptaddress]{revtex4-1}
\usepackage{graphicx}
\usepackage{dcolumn}
\usepackage{bm}
\usepackage{amsmath}
\usepackage{amssymb}
\usepackage{colortbl}
\usepackage{hhline}

\definecolor{dunkelgrau}{rgb}{0.8,0.8,0.8}
\definecolor{mittelgrau}{rgb}{0.9,0.9,0.9}
\definecolor{hellgrau}{rgb}{0.95,0.95,0.95}

\begin{document}

\title{Protein Interaction Networks are Fragile against Random Attacks and Robust against Malicious Attacks}

\author{Christian M. Schneider}
\email{schnechr@ethz.ch} \affiliation{Computational Physics, IfB, ETH
Zurich, Schafmattstrasse 6, 8093 Zurich, Switzerland}
\author{Roberto F.S. Andrade}
\affiliation{Instituto de F\'{\i}sica, Universidade Federal da Bahia,
40210-210, Salvador, Brazil}
\author{Troy Shinbrot}
\affiliation{Biomedical Engineering, Rutgers University, Piscataway, NJ, USA}
\author{Hans J. Herrmann}
\affiliation{Computational Physics, IfB, ETH Zurich, Schafmattstrasse 6,
8093 Zurich, Switzerland} \affiliation{Departamento de F\'{\i}sica,
Universidade Federal do Cear\'a, 60451-970 Fortaleza, Cear\'a, Brazil}

\date{\today}

\begin{abstract}
The capacity to resist attacks from the environment is crucial to the survival of all organisms.  We quantitatively analyze the susceptibility of protein interaction networks of numerous organisms to random and malicious attacks.  We find for all organisms studied that random rewiring improves protein network robustness, so that actual networks are more fragile than rewired surrogates. This unexpected fragility contrasts with the behavior of networks such as the Internet, whose robustness decreases with random rewiring.  We trace this surprising effect to the modular structure of protein networks.
\end{abstract}


\maketitle


\section{Introduction}

\begin{figure}
\includegraphics[width=8.cm,angle = 0]{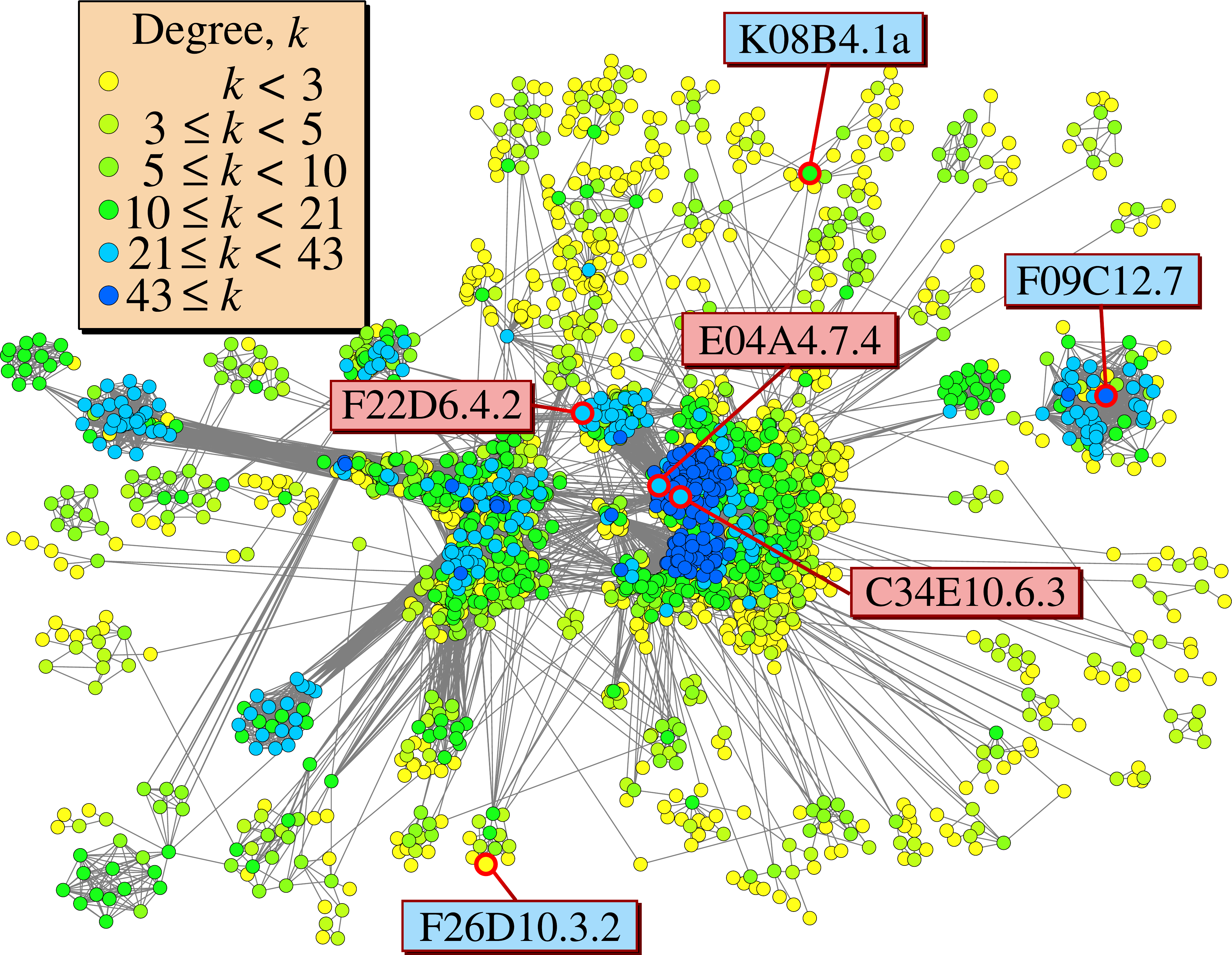}
\caption{(color online) protein network for \emph{C. Elegans} \cite{Pajek}. Modular proteins identified include F09C12.7, an element of major sperm protein, K08B4.1a involved in embryonic development and notch, and F26D10.3.2, which is a heat shock protein. On the other hand, the proteins identified in the central complex are essential to the Krebs cycle: F22D6.4.2 encodes a subunit of NADH dehydrogenase, E04A4.7.4 is better known as cytochrome c 2.1, and C34E10.6.3 is ATP synthase.} \label{fig1}
\end{figure}

Over the past two decades, prodigiously detailed maps of protein interaction networks (PPI) have been produced \cite{Kohn99,Kohn06}. These networks in principle present a record of all metabolic processes and their inter-relations, but in practice the number of chemical actors and the complexity of their interactions make the networks difficult to decipher \cite{Alon06}.  In this letter, we show that notwithstanding their apparent complexity, it is possible to establish common features of protein networks starting from a few simple principles\cite{Tononi99,Jeong01,Samanta03,Sharan07}.\\
We begin with the observation, illustrated in Fig. \ref{fig1}, that biological protein networks involve both common processes that all cells must use (e.g. enzymes involved in the Krebs cycle, marked with red labels in Fig. 1) and what are termed modular processes \cite{Hartwell99,Rives03,Ravasz02} that appear only in special situations (e.g. guidance molecules used only during particular circumstances, as development, reproduction, or response to heat stress, indicated by blue labels in Fig. 1). As we will show, this modular organization produces common, and predictable, network properties shared by all organisms studied. We focus in particular on the fragility of biological networks - a property of manifest importance for survival - to attacks by interruption of individual protein function. To this end, we evaluate the extent to which protein interaction networks of 20 different organisms ranging from bacteria and plants to homo sapiens (Table \ref{tab1}) can be disrupted by either random or targeted attacks.\\
As we have remarked, the modular construction shown in Fig. \ref{fig1} consists of a highly interconnected core of proteins, accompanied by satellite clusters with "hub" proteins weakly connected to the core. As a consequence, three predictions can readily be made. First, this type of network can be expected to be vulnerable to attacks that interrupt the few hub proteins, but should be comparatively robust against attacks that interrupt any of the more numerous proteins attached to 'spokes' of these hubs \cite{Rives03}. Thus random attacks are unlikely to significantly interrupt function, while malicious attacks directed against one or more hub proteins are likely to disrupt the network. Second, through countless generations of attacks, we expect evolution to have tuned biological networks to be more robust against attacks than statistically comparable, but non-biological, networks. Third, through the same reasoning we expect biological networks to be optimal in that alternative interconnections should worsen robustness.  As we will show, these predictions are largely correct, but admit unexpected and revealing failures.\\
\begin{figure}
\includegraphics[width=6.cm,angle = -90]{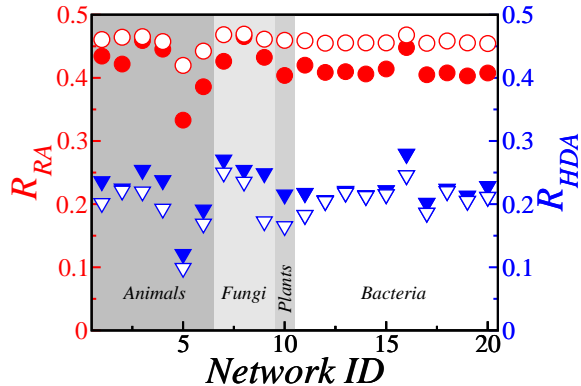}
\caption{Robustness of the 20 protein networks from Table I against random and malicious attacks.  Notice that for every organism  studied, the robustness against random attacks is \underline{smaller} than surrogates with identical degree distributions, while the robustness against malicious attacks is \underline{larger} than such surrogates. Solid and open symbols correspond, respectively, to biological data and surrogates. Error bars, defining standard deviations over 20 randomized surrogate trials, are smaller than the symbol sizes.}
\label{fig2}
\end{figure}
\begin{table}
 \begin{tabular}{l|c|c|c|c}
 Organism & ID & $N$ & $M$ & $\langle k \rangle$\\\hhline{-----}\hhline{-----} \rowcolor[gray]{.75}[\tabcolsep]
Drosophila melanogaster - AA & 1 & 3960 & 44409 & 22.4\\ \rowcolor[gray]{.75}[\tabcolsep]
Gallus gallus - AC & 2 & 3723 & 54131 & 29.1\\ \rowcolor[gray]{.75}[\tabcolsep]
Homo sapiens - AC & 3 & 12299 & 176316 & 28.7\\ \rowcolor[gray]{.75}[\tabcolsep]
Mus musculus - AC & 4 & 9595 & 123665 & 25.8\\ \rowcolor[gray]{.75}[\tabcolsep]
Xenopus tropicalis - AC & 5 & 1870 & 7374 & 7.9\\ \rowcolor[gray]{.75}[\tabcolsep]
Caenorhabditis elegans - AN & 6 & 2113 & 14261 & 13.5\\ \rowcolor[gray]{.9}[\tabcolsep]
Aspergillus fumigatus - FA & 7 & 2364 & 29288 & 24.8\\ \rowcolor[gray]{.9}[\tabcolsep]
Saccharomyces cerevisae - FA & 8 & 5209 & 66057 & 25.4\\ \rowcolor[gray]{.9}[\tabcolsep]
Schizosaccharomyces pombe - FA & 9 & 2458 & 28822 & 23.5\\ \rowcolor[gray]{.825}[\tabcolsep]
Arabidopsis thaliana - PM & 10 & 4205 & 81957 & 40.0\\
Rhodococcus sp - AB & 11 & 5540 & 57992 & 20.9\\ 
Saccharopolyspora erythraea - AB & 12 & 3715 & 24691 & 13.3\\
Aeromonas hydrophila - PB & 13 & 2765 & 13849 & 10.0\\
Bradyrhizobium japonicum - PB & 14 & 4948 & 29628 & 12.0\\
Citrobacter koseri - PB & 15 & 3477 & 17288 & 9.9\\
Escherichia coli - PB & 16 & 3542 & 25197 & 14.2\\
Nocardia farcinica - PB & 17 & 3277 & 21359 & 13.0\\
Pseudomonas aeruginosa - PB & 18 & 3709 & 20401 & 11.0\\
Serratia proteamaculans - PB & 19 & 3392 & 16978 & 10.0\\
Vibrio cholerae - PB & 20 & 2506 & 12899 & 10.3\\
\end{tabular}
\caption{List of organisms investigated. Acronyms in column 1 indicate kingdom and phylum the organisms belong to: AA - Animalia Arthropoda; AB - Actinum Bacteria; AC - Animalia Chordata; AN - Animalia Nematoda; FA - Funghi Ascomycota; PB - Bacteria Proteo; PM - Plantae Magnoliophyta. The ID (second column) is used to identify organisms in Figs. \ref{fig2} and \ref{fig4}. Columns 3, 4 and 5 define the numbers of nodes in the largest cluster $N$, total numbers of edges $M$, and average degree $\langle k \rangle$.
Shadings correspond to Fig. 2.}
\label{tab1}
\end{table}
\section{Results and discussion}
\begin{figure}
\includegraphics[width=4.6cm,angle = -90]{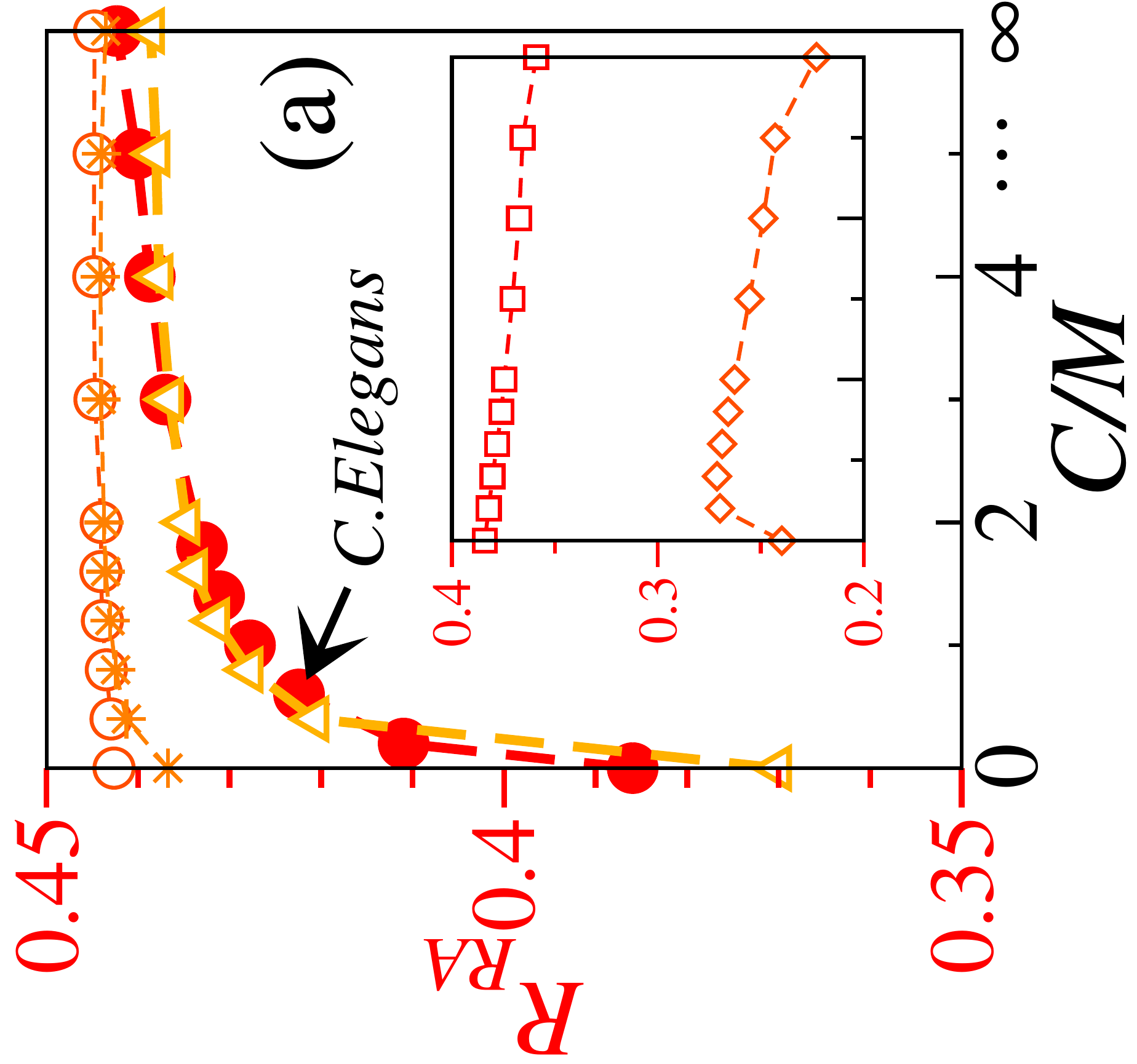}
\includegraphics[width=4.6cm,angle = -90]{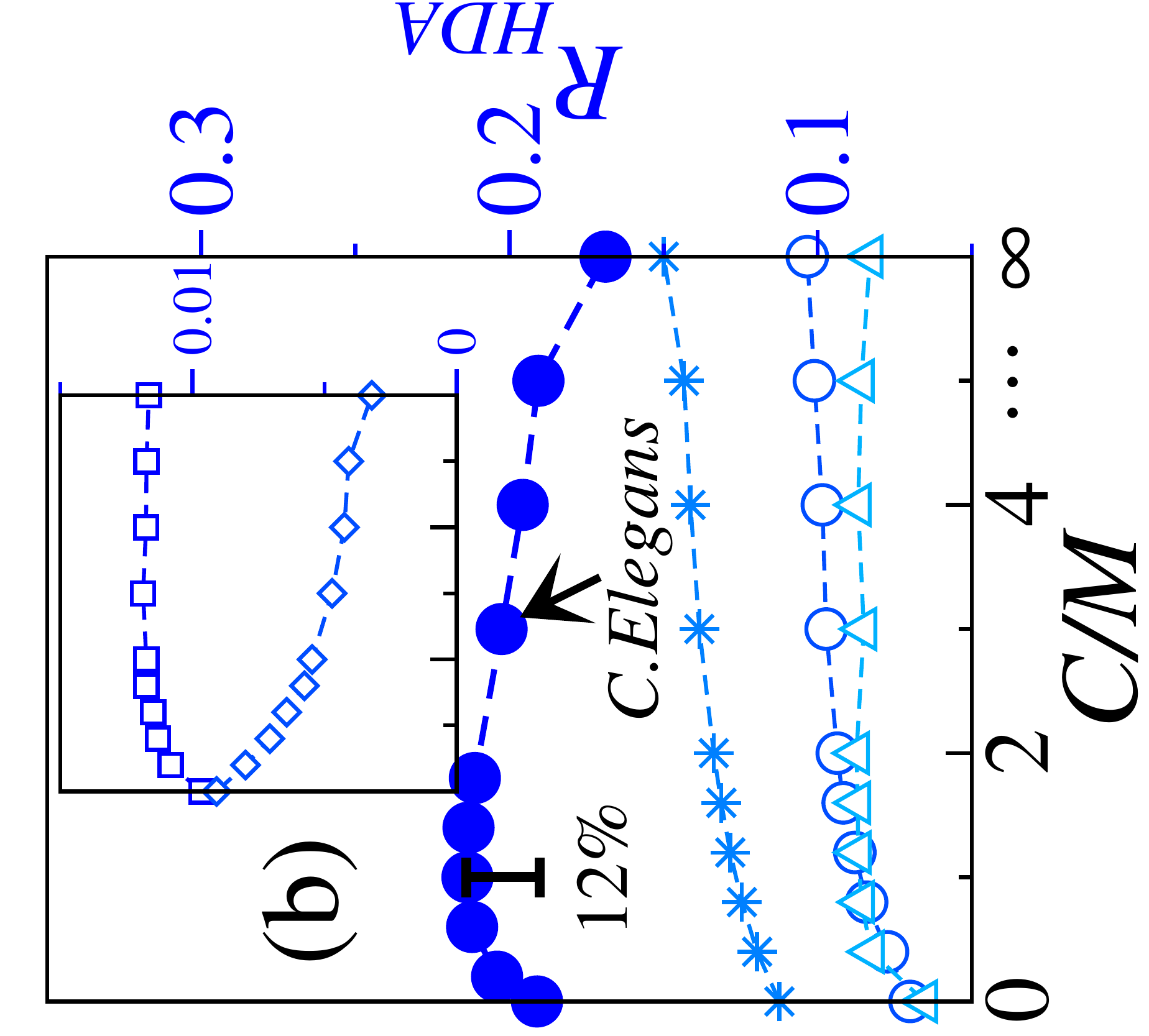}
\caption{Fundamentally different behaviors of fragile and robust networks. Robustness against random attacks, $R_\mathrm{RA}$, \underline{increases} with an increasing fraction $C/M$ of changed edges for \emph{C. Elegans}(filled circles) and other networks such as airline (triangles)\cite{Airport}, citation(stars)\cite{Citation} and point-of-presence networks (open circles)\cite{PoP}, while by contrast network robustness \underline{decreases} with $C/M$ for the Internet (squares)\cite{Dimes} and corporate ownership network (diamonds)\cite{EVA}. Note that the improvement in robustness against random attacks is significantly larger for \emph{C. Elegans} and airline networks, both of which are modular, and is opposite to that of the Internet (inset). Likewise the robustness against malicious attacks, $R_\mathrm{HDA}$, differs between biological and other networks. $R_\mathrm{HDA}$ increases with $C/M$ up to $12\%$ until $C/M \approx 1$, after which $R_\mathrm{HDA}$ decreases for biological networks, in contrast with all other networks except for the ownership network, for which $R_\mathrm{HDA}$ monotonically increases with $C/M$. For better visibility some data are shown in the insets having abscissas using the same axes as the main plot; curve fits are included to aid the eye.}
\label{fig3}
\end{figure}

\begin{figure}
\includegraphics[width=6.cm,angle = -90]{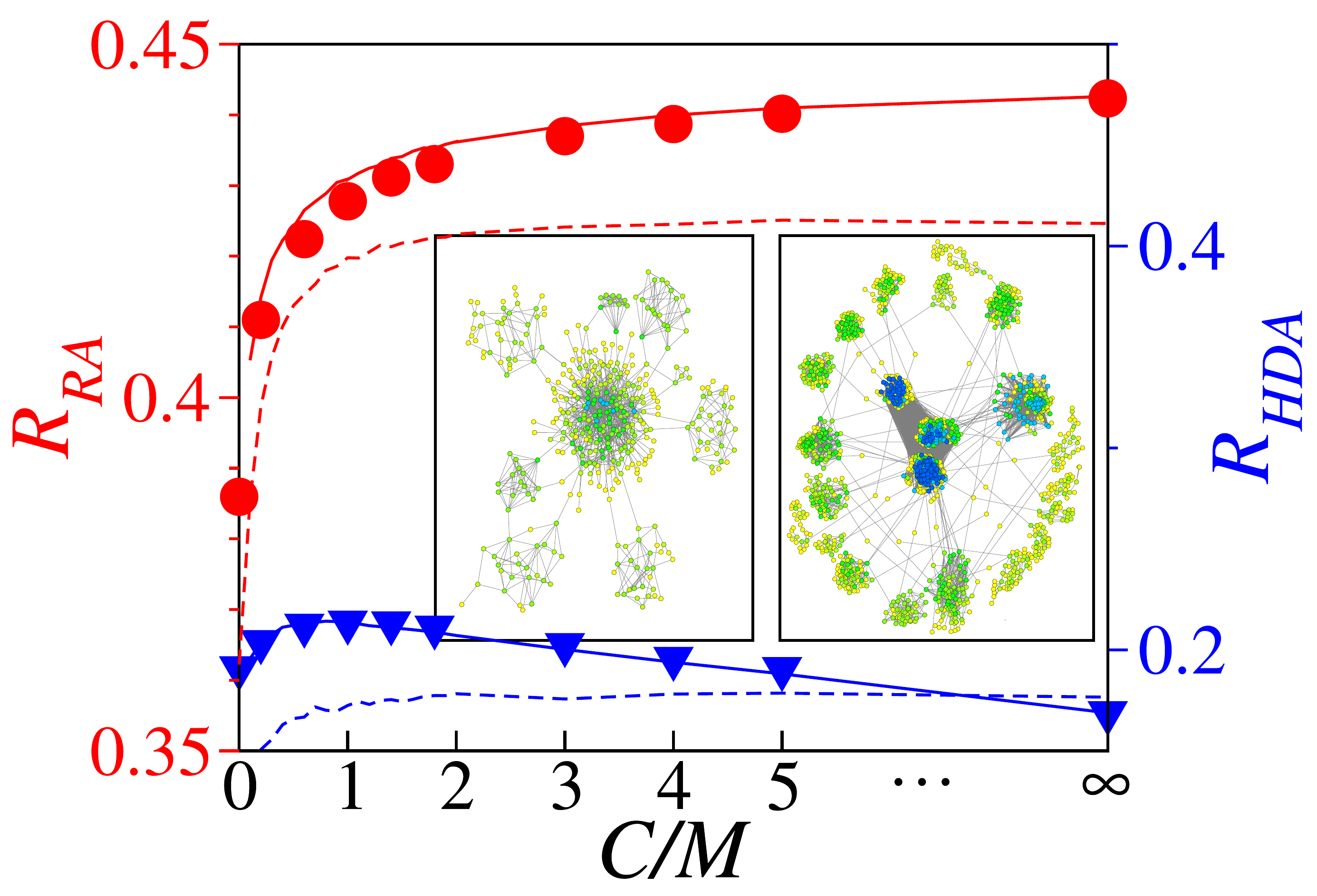}
\caption{Effects of modularity on robustness in model (curves) and sample biological networks (data points, for {\it C. Elegans}). Insets show simple network (left) and more complex network (right) fit to {\it C. Elegans} data. Network robustnesses are shown as dashed and solid lines respectively.}
\label{fig4}
\end{figure}

We examine protein networks of 20 different organisms in the bacteria and eukarya domains, identified in Table 1 along with numbers of nodes $N$ (i.e. proteins), edges $M$ (connections between proteins), and the average degree $\langle k \rangle$ (number of connections per protein) of the largest connected component of each network.
Our measure of robustness is essentially unaffected by the small number of isolated nodes that are detached from the largest cluster, so we neglect these in our analysis. We used the STRING 8.2 ''Combined Score`` (CS) \cite{string}, a measure of the likelihood that two proteins interact in a given network, to impose the criterion that edges $e_{i,j}$ are included in the network only if CS$_{ij}$ is over a threshold value, CS$_{th} = 70\%$. Smaller values of CS$_{th}$ produce dramatic growth in numbers of edges, masking relevant information with extraneous information, while larger CS$_{th}$ excludes known protein interactions \cite{string1}.\\
Typical results are presented in Figs.\ref{fig2}-\ref{fig4}, showing the dependence of robustness on random and malicious attacks for several network types. As one would expect, for all networks the tolerance to random attacks is high (Fig. 2, red data) and to malicious attacks is low (Fig. 2, blue data). However unexpectedly we find that all biological networks studied have a significantly \underline{lower} resistance to random attacks, and significantly \underline{higher} resistance to malicious attacks than do surrogates, randomized $T_M = 10^8$ times, as described previously. This paradoxical behavior is surprising, and can be analyzed in further detail as shown in Fig. 3.  In that figure, we plot detailed responses to systematic randomization, using \emph{C. Elegans} as an exemplar, compared with several non-biological networks.\\
For all networks in Fig. 3(a), we find that small amounts of random rewiring improve network robustness to random attacks; for biological and other modular networks (for example airlines, shown as triangles in the plot), the improvement is much larger than for less obviously modular networks such as citations or access points (''points-of-presence``) to the Internet.  By contrast, the behavior of a second class of highly redundant networks, for example the entire Internet or corporate ownership networks, is shown in the insets to Fig. 3. These networks are nearly optimally robust, since switching connections tends to \underline{reduce} network robustness. \\
Thus a first and unexpected finding of this analysis so far is that as shown in Fig. 3(a), biological proteins and other modular networks are less than ideally organized from the point of view of robustness against random attacks, insofar as this robustness can be significantly improved by any amount of rewiring.  A second unexpected finding, shown in Fig. 3(b), is that although biological protein networks are more than twice as robust against malicious attacks as any other network tested, modest modifications of the protein interaction structure can improve the network robustness from $2 \%$ for \emph{H. Sapiens} and $12 \%$ for \emph{C. Elegans} (Fig. 3(b)) up to $28 \%$ for \emph{G. Gallus}.
Apparently, despite the manifest two-fold improvement in robustness shown in Fig. 3(b) that evolution has produced, life remains among a class of networks that are more fragile to either random or malicious attacks than slightly modified surrogates.\\
This effect, which holds for all of the 20 organisms studied, differs markedly from a second class of networks, shown in the insets to Fig's 3(a)-(b), that is exemplified by the Internet \cite{Dimes}, which was designed for maximal robustness against errors \cite{RobustInternet}, and to a lesser extent corporate ownership networks,  that are robust by virtue of similarly numerous inter-relations \cite{EVA}.
Our findings therefore indicate that although the Internet and PPI networks share broad degree distributions, the two types of networks behave fundamentally differently in their overall fragility as measured by comparison with modified surrogates.
\\
To investigate the consistency of these results, we repeated our analyses under various modifications. First, we evaluated the reliability of the data itself by considering both a higher value of the threshold likelihood of protein interactions, CS$_{th}=80\%$, as well as data from a different version of STRING 8.1 \cite{stringold}. Second, we considered whether the robustness could be an indirect effect of a change in correlation between nodes - for example as high degree nodes are swapped with low degree ones. For this purpose, we modified the rewiring to preserve correlations by performing swaps between pairs of nodes $\{(i,j),(k,\ell)\} \rightarrow \{(i,\ell),(k,j)\}$ only if the degrees of $i$ and $k$ or $j$ and $\ell$ are equal.  
Third, we considered the effect of randomly removing individual edges described again by Eq. \ref{eq2}, but with $N$ defined to be the number or edges, rather than nodes. In each of these independent tests, we found the same features commented on for Fig's \ref{fig2} and \ref{fig3}, supporting the two key results that biological networks exhibit more fragility to random errors than similar non-biological networks, and that although biological networks are more than twice as robust against malicious attacks as non-biological networks, they remain less than optimal.\\
We have reported three curious and previously unexplored properties of protein-protein interaction (PPI) networks.  (1) PPI networks are {\it less} robust against random attacks than surrogates with identical numbers of nodes, edges, and degree distributions; (2) PPI networks are {\it more} robust against malicious attacks than surrogates; and (3) despite millenia of evolutionary pressure, PPI network robustness against malicious attacks remains suboptimal and can be improved by modest rewiring.

To analyze the causes of these unexpected behaviors, we return to the observation made earlier that PPI networks are intrinsically modular.  Since modules have many fewer nodes than the central network, it follows that any switch involving a node in a module is highly likely to involve a second connection that is outside of the module.  Such a switch will produce two new edges, both of which will connect the module to the central network, so switching connections will typically increase the number of connections from a module to the central network. This in turn will improve the robustness of that module to either random or malicious attack, since such a switch would increase the number of connections that would have to be broken between the module and the rest of the network.

We can test this mechanism by constructing model networks with suitable properties, two of which are shown in Fig. 4.  In that figure, we compare both simple and more finely tuned model networks with the biological data that appear as solid symbols in Fig. 3. The simple model (left inset) is constructed by creating a central large complex with broad degree distribution. An arbitrary small number (eight, here) of modules, each with different number of nodes but the same number of random connections, are added and are attached to random nodes in the central complex by two connections. The robustness in response to random and malicious attacks is then evaluated exactly as before, and is plotted in Fig. 4 as dashed curves. By contrast, non-modular model networks that we have constructed (not shown) have few vital hubs and so exhibit identical responses to either random or malicious attacks, with no dependence on $C/M$.

Evidently, the qualitative behavior of biological responses to random as well as malicious attacks can be attributed to the modular structure of biological protein networks. Indeed, it is not difficult to tune the model network to nearly exactly fit the biological data. This is shown in the right inset of Fig. 4, where we display a fictitious network whose random and malicious response curves are shown as solid lines in the main plot. This network is constructed by choosing the number of connections of the model to be similar to the biological one. In detail, the nodes are distributed in 20 modules with different densities, in which high degree nodes are preferentially connected to high degree nodes. This preferential connectivity is crucial to the reduction in robustness to malicious attacks: for no other structural feature investigated was this reduction seen. These modules are connected preferentially to the largest module with few connections, as we have remarked occurs in biological networks.

Thus the first, simpler, network of Fig. 4 demonstrates that the presence of modularity is sufficient to qualitative account for most of the curious behaviors of PPI network robustness in response to both random and malicious attacks.


Back to Fig's 2 and 3, they show that for surrogates with large numbers of switches of connections, the robustness of PPI networks to malicious attacks actually \underline{decreases} for all organisms studied. This behavior can also be reproduced in model networks provided, crucially, that connections are preferentially included between high degree nodes. In this case, two competing effects arise. The randomization of the modules increases robustness, while the vanishing of the preferential connections decreases the robustness. In case of random attack the second effect is negligible, but for malicious attacks, it leads to the surprising decrease in robustness that we have noted.


In conclusion, we have demonstrated that biological protein networks are unexpectedly fragile against either random or malicious attacks.  This fragility is measurable by comparison with surrogates with identical network statistics.  We find that these behaviors are characteristic of modular networks, in which particular products or processes inhabit isolated modules.  As anticipated earlier in this letter, we have confirmed (1) that this modular structure causes biological protein networks to be more vulnerable to targeted than random attacks, and (2) that through evolution these networks have become more robust than non-biological networks against malicious attacks.  Nevertheless, as we have shown, protein networks are more fragile than extensively rewired surrogates to random attacks, while being \underline{less} fragile than the same surrogates to malicious attacks.  We find that this final phenomenon is associated with the apparently unique tendency of high degree nodes in PPI networks to preferentially connect to other high degree nodes.  We speculate that this preferential connectivity may have practical advantages, for example in providing redundant pathways to permit key processes to function after a malicious attack or genetic deletion \cite{Tononi99,Jeong01,Samanta03}.\\

\section{Methods}
To test these predictions, we compare known protein networks with surrogates that are as statistically similar as possible. To this end, we generate randomized surrogate networks having the same size and degree distribution as true biological networks. To create such surrogates, we perform a sequence of randomly chosen switches of connections between pairs of nodes $\{(i,j),(k,\ell)\} \rightarrow \{(i,\ell),(k,j)\}$ in a network, so that each node preserves its number of neighbors \cite{Maslov02}.
The randomizing algorithm is repeated $T_M$ times, where $T_M$ ranges from $0$ to $10^8$. For the organisms we study, $T_M = 10^8$ ensures that each edge has been swapped more than $10^3$ times, effectively destroying any initial correlation in the network. We evaluate correlations between nodes by calculating nearest neighbor average connectivity\cite{Pastor01}
\begin{eqnarray}
 \overline{k}_{nn} (k) = \sum_{k'}k' P(k'|k),
\end{eqnarray}
where $P(k'|k)$ is the conditional probability that a node with degree $k$ is connected with one of degree $k'$. Indeed the created surrogates are uncorrelated.\\
Given a network and its surrogates, we evaluate the "robustness" (defined shortly) of the network to random or targeted, malicious, attacks. For biological networks, random attacks (RA)\cite{Albert00} take into account single gene changes due to radiation or mutagen exposure and errors in transcription. By contrast, malicious attacks describe situations in which pathogens or toxins interfere with high degree hubs of the network. Such an attack is termed a "high degree based adaptive attack (HDA)" in the literature \cite{Holme02,Cohen00,Tanizawa05}.
To define the robustness of a network against either random or malicious attack, we evaluate the sum of the fractions of the largest connected cluster while removing all nodes,
\begin{eqnarray}
R = \frac{1}{N + 1} \sum_{q = 0}^{N} s(q),
\label{eq2}
\end{eqnarray}
where $N$ is the number of nodes in the network and $s(q)$ is the fraction of nodes in the largest connected cluster after removing $q$ nodes. This measure has the advantage over other, e.g. percolation \cite{Holme02}, metrics of robustness in that it can distinguish between different networks with similar ''percolation thresholds``, at which a significant number of elements of a network form a single cluster \cite{Schneider10}.
The normalization $\frac{1}{N+1}$ in Eq. (\ref{eq2}) ensures that the robustness is comparable for different network sizes, and the value of $R$ lies between $\frac{1}{N+1}$ and $0.5$. The lower limit on $R$ corresponds to entirely isolated nodes, and $R=0.5$ defines a network where all unattacked nodes remain in a single cluster.\\

\section{Acknowledgment}
The authors thank Prof. R.M.C. Almeida for helpful discussions on the use of biological data. C.M. Schneider and H.J. Herrmann acknowledge financial support from the ETH Competence Center 'Coping with Crises in Complex Socio-Economic Systems' (CCSS) through ETH Research Grant CH1-01-08-2 and FUNCAP. R.F.S. Andrade thanks CNPq, SECTI-FAPESB-PRONEX, and INCTI-SC for financial support.

\bibliographystyle{prsty}

\begin{thebibliography}{10}


\bibitem{Kohn99} Kohn KW (1999) Molecular interaction map of the mammalian cell cycle control and DNA repair systems. {\it Molec. Biol. Cell} {\bf 10 (8)}, 2703-2734

\bibitem{Kohn06} Kohn KW, Aladjem MI, Kim S, Weinstein JN, Pommier Y (2006) Depicting combinatorial complexity with the molecular interaction map notation. {\it Mol. Sys. Biol.} {\bf 51}, 1-12

\bibitem{Alon06} Alon U (2003) Biological networks: The tinkerer as an engineer. {\it Science} {\bf 301}, 1866-1867

\bibitem{Tononi99} Tononi G, Sporns O, Edelmann GM (1999) Measures of degeneracy and redundancy in biological networks. {\it PNAS} {\bf 96}, 3257-3262

\bibitem{Jeong01} Jeong H, Mason SP, Barab\'asi AL, Oltvai ZN (2001) Lethality and centrality in protein networks. {\it Nature} {\bf 411}, 41-42

\bibitem{Samanta03} Samanta MP, Liang S (2003) Predicting protein functions from redundancies in large-scale protein interaction networks. {\it PNAS} {\bf 100}, 12579-12583

\bibitem{Sharan07} Sharan R, Ulitsky I, Shamir R (2007) Network-based prediction of protein function. {\it Mol. Sys. Biol.} {\bf 3}, 88

\bibitem{Ravasz02} Ravasz E, Somera AL, Mongru DA, Oltvai ZN, Barab\'asi AL (2002) Hierarchical organization of modularity in metabolic networks. {\it Science} {\bf 297}, 1551-1555

\bibitem{Rives03} Rives AW, Galitski T (2002) Modular organization of cellular networks. {\it PNAS} {\bf 100}, 1128-1133

\bibitem{Hartwell99} Hartwell LH, Hopfield JJ, Leibler S, Murray AW (1999) From molecular to modular cell biology. {\it Nature} {\bf 402}, C42-C47

\bibitem{Maslov02} Maslov S, Sneppen K (2002) Specificity and stability in topology of proteins networks. {\it Science} {\bf 296}, 910-913

\bibitem{Pastor01} Pastor-Satorras R, V\'azquez A, Vespignani A (2001) Dynamical and correlation properties of the Internet. {\it Phys. Rev. Lett.} {\bf 87}, 258701

\bibitem{Albert00} Albert R, Jeong H and Barab\'asi AL (2000) Error and attack tolerance of complex networks. {\it Nature} {\bf 406}, 378

\bibitem{Holme02} Holme P, Kim BJ, Yoon CN, Han SK (2002) Attack vulnerability of complex networks. {\it Phys. Rev. E} {\bf 65}, 056109

\bibitem{Cohen00} Cohen R, Erez K, ben-Avraham D, Havlin S (2000) Resilience of the Internet to random breakdowns. {\it Phys. Rev. Lett.} \textbf{85}, 4626

\bibitem{Tanizawa05} Tanizawa T, Paul G, Cohen R, Havlin S, Stanley HE (2005) Optimization of network robustness to waves of targeted and random attacks. {\it Phys. Rev. E} {\bf 71}, 047101

\bibitem{Schneider10} Schneider CM, Moreira AA, Andrade Jr. JS, Havlin S, HJ Herrmann (2010) Mitigation of malicious attacks on networks. {\it PNAS} {\bf }

\bibitem{string} http://string-db.org

\bibitem{string1} von Mering C {\it et al.} (2005) STRING: known and predicted protein–protein associations, integrated and transferred across organisms. {\it Nucleic Acids Research} {\bf 33}, D433-D437

\bibitem{Dimes} www.netdimes.org.

\bibitem{RobustInternet} Willinger W, Doyle J (2002) Robustness and the Internet: Design and evolution. In {\it Robust design: A Repertoire of Biological, Ecological, and Engineering Case Studies}, Jen E (ed), Oxford University Press, New York

\bibitem{EVA} Norlen K, Lucas G, Gebbie M, Chuang J (2002) EVA: Extraction, visualization and analysis of the telecommunications and media ownership network. {\it Proceedings of International Telecommunications Society 14th Biennial Conference}

\bibitem{stringold} http://string-db.org/server\_versions.html

\bibitem{Pajek} Batageli V, Mrvar A (2008) {\it http://vlado.fmf.uni-lj.si/pub/networks/pajek/} {\bf V 1.23}

\bibitem{Airport} Colizza V, Pastor-Satorras R, Vespignani A (2007) Reaction-diffusion processes and metapopulation models in heterogeneous networks. {\it Nature Physics} {\bf 3}, 276-282

\bibitem{Citation}
http://vlado.fmf.uni-lj.si/pub/networks/data/cite/default.htm

\bibitem{PoP}
Shavitt Y, Zilberman N (2010) A structural approach for PoP geo-location. {\it NetSciCom}

\end{thebibliography}

\end{document}